\documentstyle[11pt]{article}

\begin{document}
\begin{flushright}
CUMT-MATH-9504\par
\end{flushright}
\vspace{.5cm}
\centerline{\bf\Large{\bf{Painlev\'{e} Analysis and Exact Solutions
}}}
\bigskip
\centerline{\bf\Large{\bf{of a Modified Boussinesq Equation}}}
\vspace{1in}

\centerline{\large Q.\ P. \ Liu}
\smallskip
\smallskip
\centerline{CCAST(World Laboratory), P.O.Box 8730, Beijing 100080, China.}
\smallskip
\smallskip
\centerline{\bf\large and}
\smallskip
\smallskip
\centerline{Beijing Graduate School(\#65),}
\centerline{China University of Mining and Technology, }
\centerline{ Beijing 100083, China\footnote{Mailing address}}

\newcommand{\be}{\begin{equation}}
\newcommand{\ee}{\end{equation}}
\newcommand{\ba}{\begin{array}}
\newcommand{\ea}{\end{array}}
\def\w42{$W^{(2)}_4$}
\def\p{\partial}
\def\alf{\alpha}
\def\bi{\beta}
\def\es{\epsilon}
\def\la{\lambda}
\def\dl{\delta}
\vspace{0.5in}\begin{center}
\begin{minipage}{5in}
{\bf ABSTRACT}\hspace{.2in}We consider a modified Boussinesq type equation.
The Painlev\'{e} test of the WTC method is performed for this
equation and it shows that the equation has weak Painlev\'{e} property.
Some exact solutions are constructed.
\par
\end{minipage}

\end{center}
\vspace{.5in}
\vfill\eject
\par
Great achievements have been made for the Soliton theory
during last three decades\cite{ab}. Now there exist several
methods to test integrability for a given equation.
Among them, Painlev\'{e} analysis
of Weiss, Tabor and Carnevale(WTC)\cite{wis1} are particularly
interesting and has seen a great success. Applying the WTC
Painlev\'{e}
test to integrable systems, one may extract some important
information. For example, Lax pairs, B\"{a}cklund transformations
may be rediscovered this way\cite{wis2}\cite{cot1}. \par

It was found that the WTC method may also be applied to nonintegrable
systems. Though these systems may not pass the test in full
sense, it may have weak Painlev\'{e} property. More importantly,
some useful results can be produced from this test, such as
exact solutions, B\"{a}cklund transformations. So far, several
nonintegrable equations are considered and we here just cite
the n-dimensional Sine-Gordon equation\cite{wis3}, the Kuramoto-
Sivashinsky equation\cite{cot2}, the real Newell-Whitehead and
the Ginsburg-Landau equations\cite{tab}, and the carrier flow
equation\cite{gu}(see \cite{clark} for more references).\par
In this note, we give one more example. The equation we will
consider is
\be
u_{tt}=(u_{xx}-u^3)_{xx}
\label{p}
\ee
this is a real counterpart of the equation derived by Paldor\cite{pd}.
Indeed, Paldor proposed the following equation
$$
\hat{u}_{tt}=(\hat{u}_{xx}+\hat{u}^3)_{xx}
$$
as a model to describe nonlinear waves on a coupled
density front. He further considered the existence of
solitary wave solution by means of qualitative
analysis. However, an explicit solitary solution is not
given. In fact, it is easy to see that following solution
$$
\hat{u} =\pm \sqrt{2}k sech(kx\pm k^2t+\delta)
$$
is the solitary solution Paldor meant\cite{pd}.
\par
Compared with Boussinesq equation, we recognize that this system(1)
has higher nonlinear term - cubic rather quadratic. This
reminds of the relation of the KdV and modified
KdV equations. That is in this sense, the equation is modified
Boussinesq type. \par

Let us rewrite the equation(1) as a two-component system
\be
g_t=h_x, \quad \quad ~h_t=(g_{xx}-g^3)_x
\ee
where we used $g\equiv u$. Now it is interesting that the equation
(2) is Hamiltonian
\be
\left[\ba{cc}g\\[1mm] h\ea\right]_t=
\left[\ba{cc} 0&\p\\[1mm] \p&0 \ea\right]
\left[\ba{cc}{\delta{H_{1}}\over\delta g}\\[1mm]
{\delta{H_{1}}\over\delta h}
\ea\right]
\ee
where $H_1=-{1\over 2}g_{x}^{2}-{1\over 4}g^4 +{1\over2}h^2$. The other
more obvious conserved quantities are $H_{-2}=g$, $H_{-1}=h$ and
$H_0=gh$.
\par
Before doing Painlev\'{e} analysis, Let us note that the following
solitary wave solution exists for the equation(1)
\be
u=\pm \sqrt{2}k csch(kx\pm k^2t+\delta)
\ee
where k and $\delta$ are constants.

We will see that the system is weak Painlev\'{e}. Interestingly, the
compatible conditions required lead to the well-known inviscid Burgers
equation(or Hopf equation). Some exact solutions are constructed.
By truncating the Laurent series, a B\"{a}cklund transformation is
produced.\par

As known, we take the following Laurent expansion of the function
u(x,t) about a singular manifold $\phi(x,t)$
\be
 u(x,t)=\phi^p\sum_{j=0}^{\infty}u_j\phi^j
\ee
where $u_j=u_j(x,t)$. By means of leading order analysis,
we find that leading power is $p=-1$. Substituting above
expansion with  $p=-1$ to our equation and equalizing the
coefficient of the $\phi^{j-5}$ to zero,
we find that
\be
P_1=P_2-6P_3-3P_4 \label{exp}
\ee
where $P_i,i=1,...,4$ are given by

\bigskip

\par
\begin{tabular}{ll}
$P_1 =$ & $(j-4)((j-3)\phi_{t}^{2}u_{j-2} + \phi_{tt}u_{j-3}
	   +2\phi_t u_{j-3,t}) + u_{j-4,tt}$,\\[1.2mm]
$P_2 =$ & $(j-1)(j-2)(j-3)(j-4)\phi_{x}^4 u_j +
	  (j-2)(j-3)(j-4)(6\phi_{x}^{2}\phi_{xx}u_{j-1}+$  \\[1mm]
	& $4\phi_{x}^{3}u_{j-1,x})+(j-3)(j-4)(3\phi_{xx}^{2}u_{j-2} +
	  4\phi_x\phi_{xxx}u_{j-2}+12\phi_x\phi_{xx}u_{j-2,x}+$\\[1mm]
	& $6\phi_{x}^2 u_{j-2,xx})+(j-4)(\phi_{xxxx}u_{j-3} +
	   4\phi_{xxx} u_{j-3,x}+
	   6\phi_{xx}u_{j-3,xx} +4\phi_x u_{j-3,xxx})+$\\[1mm]
	& $u_{j-4,xxxx}$,\\[1.2mm]
$P_3 =$ & $\phi_{x}^{2}\sum_{m=0}^{j} u_{j-m}A_m+
	  \sum_{m=0}^{j-2}u_{j-2-m}B_m +
	  \phi_{x}^2\sum_{m=0}^{j}u_{j-m}C_m - $  \\[1mm]
	& $2\phi_x\sum_{m=0}^{j-1}u_{j-1-m}D_m -
      2\phi_{x}^{2}\sum_{m=0}^{j}u_{j-m}E_m +
      2\phi_x\sum_{m=0}^{j-1}u_{j-1-m}F_m,$ \\[1.2mm]
$P_4 =$ & $2\phi_{x}^{2}\sum_{m=0}^{j}A_{j-m}u_m -
	  \phi_{xx}\sum_{m=0}^{j-1}A_{j-1-m}u_m -
	  2\phi_{x}\sum_{m=0}^{j-1}A_{j-1-m}u_{m,x} -$ \\[1mm]
	& $ 2\phi_{x}^{2}\sum_{m=0}^{j}mA_{j-m}u_m +
	    \sum_{m=0}^{j-2}A_{j-2-m}u_{m,xx} +
	    2\phi_{x}\sum_{m=0}^{j-1}mA_{j-1-m}u_{m,x} + $\\[1mm]
	& $ \phi_{xx}\sum_{m=0}^{j-1}mA_{j-1-m}u_m +
	    \phi_{x}^{2}\sum_{m=0}^{j}m(m-1)A_{j-m}u_m.$
\end{tabular}

\bigskip

where $A_j$, $B_j$, $C_j$ $D_j$,$ E_j$ and $F_j$ are defined by

\bigskip
\par
\begin{tabular}{lll}
$A_j  = \sum_{k=0}^{j}u_{j-k}u_k,$
 & $B_j = \sum_{k=0}^{j}u_{j-k,x}u_{k,x},$
 & $C_j = \sum_{k=0}^{j}(j-k)ku_{j-k}u_k,$ \\[1.5mm]
$ D_j  =  \sum_{k=0}^{j}u_{j-k}u_{k,x},$
 & $E_j = \sum_{k=0}^{j}ku_{j-k}u_k,     $
 & $F_j =  \sum_{k=0}^{j}ku_{j-k,x}u_k $
\end{tabular}

\bigskip

The resonances are found from the roots of the following
equation
\be
(j+1)(j-3)(j-4)^2=0
\ee
which are $j=-1, j=3$ and $j=4,j=4$
\par
To convenience, we introduce the following notions as in\cite{tab}
\be
w={\phi_t\over\phi_x},\quad \quad~~v={\phi_{xx}\over\phi_x}
\ee
{}From the above rather long expression (\ref{exp}), we have
\be
j=0:  \quad \quad u_0=\sqrt{2} \phi_x
\ee
\be
j=1: \quad \quad u_1=-{\sqrt{2}\over2}v
\ee
\be
j=2: \quad \quad~u_2={\sqrt{2}\over6\phi_x}(-{1\over2}v^2-w^2+v_x)
\ee
\be
j=3: \quad \quad ~~~~ w_t=ww_x \label{Bg}
\ee
\be
j=4: \quad \quad~~ (w_t-ww_x)_x +v(w_t -ww_x)=0 \label{mbg}
\ee
Thus, the equation(1) does not pass the WTC Painlev\'{e}
test in the full sense since
the compatibility condition at the resonances $j=3,4$ are not
satisfied identically.

However, if we impose that the singular manifold satisfys
the equation(\ref{Bg}), then the condition(\ref{mbg})
at the resonance $j=4$ is hold consequently.
\par
\bigskip

{\em Remark}. If we choose the Kruskal gauge for the singular
manifold $\phi =x-f(t)$ at the very beginning, we obtain a
constraint at $j=3$. But at $j=4$, compatibility condition
is satisfied automatically.
\par

\bigskip

The equation(\ref{Bg}) is well-known and referred as inviscid Burgers
equation  in the literatures. Its classical theory is available from
the book\cite{whm}. The Interesting Hamiltonian properties
of the equation
is explored by Olver and Nutku\cite{olv} recently.

\bigskip

{\em Remark}. We notice that the WTC's Painlev\'{e} analysis of the
double Sine-Gordon equation\cite{wis2} also leads to the
inviscid Burgers equation, a typical
equation describing shock wave.
\par

\bigskip
Let us see what kind of solutions we may construct for the
equation(1). At this point, we see that $w_1=0$ and $w_2=b$
(b is an arbitrary constant) are solutions of the inviscid Burgers
equation(11). In the first case, $\phi=C(x)$ and we are lead to the
stationary equation:
\be
(u_{xx}-u^3)_{xx}=0
\ee
i.e.
\be
{d^2 u(x)\over dx^2}-u^3=C_0+C_1x
\ee
is integrable.
\par
In the second case: $w_2=b$, the solution of the equation $\phi_t=
b\phi_x$ is
\be
\phi(x,t)=f(x+bt)
\ee
f is an arbitrary function. This reduction is a particular
travelling wave
reduction of the equation(1)
\be
f''-f^3=b^2f+C_2+C_3\xi
\ee
where $\xi=x+bt$ and prime denotes the derivative with $\xi$.
The second order ODE is integrable. We will present a solution of this
equation later on.

Let us now impose the truncation condition for the Laurent
expansion(4):
\be
u={u_0\over\phi}+u_1
\label{trac}
\ee
then by letting the coefficients of the power $\phi$ to
zero, we have
\be
\phi^5 : \quad \quad u_0=\pm \sqrt{2} \phi_x
\ee
\be
\phi^4 : \quad \quad u_1=\mp {\sqrt{2}\over2}v
\ee
\be
\phi^3 :\quad \quad  v_x={1\over 2}v^2 +w^2  \label{Fa2}
\ee
\be
\phi^2 : \quad ~~~~w_t+3ww_x+3w^2v-vv_x +
{3\over 2}v^3 -2v_{xx}=0 \label{Fa3}
\ee
\be
\begin{tabular}{ll}
$\phi$ :& $ v_{xxx}=w_{tx}+w_{x}^{2}+ww_{xx}+3ww_xv+$\\
	& $ \quad \quad w^2v_{x}+w_tv+v^2w^2+{3\over 2}v^2v_x+
	    {1\over2}v^4-vv_{xx}$ \label{Fa4}
\end{tabular}
\ee
\be
\phi^0 :\quad \quad   u_{1,tt}=(u_{1,xx}-u_{1}^{3})_{xx}
\ee
Straightforward calculation shows that the equation(\ref{Fa3})
leads to the equation(\ref{Bg}) under the condition
(\ref{Fa2}).
The equations(\ref{Fa2}) and (\ref{Fa3}) together show that
the equation(\ref{Fa4}) is hold. The integrable
condition $\phi_{t,xx}=\phi_{xx,t}$ gives us an extra equation
for w and v: $v_t=w_{xx}+(wv)_x$. Thus, we have three
equations to solve. For clarity, we collect them below
\be
v_x={1\over2}v^2+w^2  \label{Rct}
\ee
\be
w_t=ww_x   \label{Mbg}
\ee
\be
v_t=v_{xx}+(vw)_x   \label{ht}
\ee
Any solution of this system(\ref{Rct}),(\ref{Mbg})
and (\ref{ht}) will provides us
a B\"{a}cklund transformation thorough the
equation(\ref{trac}) in principle.

Next we consider this system in some detail. Using $v_{xt}=v_{tx}$,
we have from the equations(\ref{Rct}), (\ref{ht}) and
(\ref{Mbg})

\be
w_{xx}+{2\over3} w^3=C(t)
\label{lst1}
\ee
where C(t) is an integration constant with respect to x.

Taking $w_{t,xx}=w_{xx,t}$ into consideration, we find
the C(t) must satisfy
\be
{dC(t)\over dt}=w_x(3C(t) - 2w^3)
\label{lst2}
\ee
a linear equation for C(t).
\par
We may eliminate $w_x$ from the equation(\ref{lst2})
using the equation(\ref{lst1})
\be
{dC(t)\over dt}=\pm (3C(t)-2w^3)\sqrt{-{1\over3}w^4+2C(t)w+D(t)}
\label{cw}
\ee
where D(t) arises from the integration with respect to x.

Therefore, we may solve the equation(\ref{cw}) for w and using
the Riccati type equation(\ref{Rct}) and the equation(\ref{ht})
to find v(x,t). This will give us a solution of the equation(1)
even without finding the explicit form of the singular manifold
$\phi$ . Of Course, whenas a $\phi$ is calculated out, we
may use the
truncation equation(\ref{trac}) to get more solutions by iteration.

We must stress that above procedure may still have difficulties
to follow since the Riccati type equation(\ref{Rct}) may present
problem. However, this method does give us interesting results as
we do next.
\par
In the case ${dC(t)\over dt}=0$, we get $w=b$(b is a constant).
The singular manifold is $\phi(x,t)=f(x+bt)$ and by the definition
of v or equation(\ref{ht}), we see $v(x,t)=g(x+bt)$
(g is an arbitrary function). Now the equation(\ref{Rct})
tells us that g is solution the following Riccati equation
\be
g'={1\over2}g^2+b^2
\label{riccati}
\ee
Linearizing the equation(\ref{riccati}) by $g=-2{{F'}\over F}$,
we find
\be
F''=-{b^2\over2}F
\ee
its general solution is
\be
F(\xi)= C_1 sin({\sqrt{2}b\xi\over 2}) +
	C_2 cos({\sqrt{2}b\xi\over 2})
\ee
Therefore, a solution of the equation(1) is
\be
u_1(\xi)=\mp {\sqrt{2}\over2}g=\pm b{C_1 -C_2 tan({\sqrt{2}\over2}b\xi)
\over C_1 tan({\sqrt{2}\over2}b\xi)+C_2}
\label{u1}
\ee
Consequently, the singular manifold in this case is
\be
\phi(\xi)={2\sqrt{2}C_3\over C_2 b}{ sin({\sqrt{2}b\xi\over 4})
cos({\sqrt{2}b\xi\over 4})\over 2C_1sin({\sqrt{2}b\xi\over 4})
cos({\sqrt{2}b\xi\over 4})+2C_2cos^2({\sqrt{2}b\xi\over 4})-C_2}
\label{phai}
\ee
where $C_i, i=1,2,3$ and b are constants and $\xi=x+bt$.
Substituting (\ref{u1}) and (\ref{phai}) into the
equation(\ref{trac}), we have any solution.
\par
Thus, by means of the WTC approach, we get various solutions
of the equation(1). It further shows that this method is
very fruitful even for nonintegrable systems.
\par
we conclude this paper with the remarks in order:
\begin{enumerate}
 \item The analysis above indicates that the equation(1)
 is not integrable. It will be interesting to perform a
numerical analysis for this equation to see what happens;
\item The equation is easily seen to have scaling symmetry.
It would be important to do classical Lie group or even
nonclassical group analysis for it. This may lead to
some new solutions;
\item Hirota's bilinear method is another effective way to
get exact solutions. We suppose that a bilinearization of the
equation(1) will provide two-soliton solution.
\item The similar consideration should be taken for the Paldor's
equation(2).
\end{enumerate}
\par
Some of these problems are under investigation.

\bigskip
\bigskip

{\bf{ACKNOWLEDGEMENT}}\par
I would like to thank Drs. X.B.Hu, S.Y. Lou and J. Zhu
for the helpful discussions. This work is
supported by Natural National Science
 Foundation of China.\par
\vspace{.1in}
\bigskip

\smallskip
\par


\small
\begin{thebibliography}{}

\bibitem{ab} M.J. Ablowitz and H.Segur, {\em Solitons and the
	     Inverse Scattering Transform},
	     SIAM, Philadelphia(1981).

	   ~~M.J.Ablowitz and P.A.Clarkson, {\em Solitons, Nonlinear
	     Evolution Equations and Inverse Scattering},
	     Cambridge University Press(1991).

\bibitem{wis1} J. Weiss, M. Tabor and G. Carnevale, J. Math. Phys.
	     {\bf 24} (1983) 522-6.

\bibitem{wis2} J. Weiss, J. Math. Phys. {\bf 23} (1983)1405-13.

\bibitem{wis3} J. Weiss, J. Math. Phys. {\bf 25} (1984) 2226-35.

\bibitem{cot1} M. Musette and R. Conte, J. Math. Phys.
		{\bf 32} (1991)1450-7.

\bibitem{cot2} R. Conte and M. Musette, J. Phys. A: Math.Gen.
	       {\bf 22} (1989) 169-77.

\bibitem{tab} F. Cariello and M. Tabor, Physica {\bf 39D} (1989) 77-94.

\bibitem{gu} Z. Chen and Y.Guo, J.Phys.A: Math. Gen. {\bf 22}
	   (1989) 5187-94.

\bibitem{clark} P. Clarkson and E. Mansfield, Physica {\bf 70D}
               (1994) 250-88.

\bibitem{whm} G. B. Whitham, {\em Linear and Nonlinear Waves},
	       Wiley-Interscience, New York(1974).

\bibitem{olv} P.J. Olver and Nutku, J. Math. Phys. {\bf 29}
	      (1989).

\bibitem{pd} G. Paldor, Geophys. Astrophys. Fluid Dyn. {\bf 37}
	     (1987) 171-98;

	  ~~ M. Ghil and G. Paldor, Geophys. Astrophys.
	     Fluid Dyn. {\bf 58} (1991) 225-41;

	  ~~ M. Ghil and G. Padlor, J. Nonlinear Sic. {\bf 4}
	     (1994) 471-96.

\end{thebibliography}
\end{document}